\documentclass[aps,prd,onecolumn,groupedaddress,showpacs,nofootinbib,amssymb]{revtex4}

\usepackage{graphicx}
\usepackage[english]{babel}
\usepackage{amsmath}
\usepackage{amssymb}
\usepackage{amsfonts}
\usepackage{color}

\allowdisplaybreaks[4] 

\begin{document}

\newcommand{\be}{\begin{equation}}
\newcommand{\ee}{\end{equation}}
\newcommand{\bea}{\begin{eqnarray}}
\newcommand{\eea}{\end{eqnarray}}
\newcommand{\beaa}{\begin{eqnarray*}}
\newcommand{\eeaa}{\end{eqnarray*}}
\newcommand{\Lhat}{\widehat{\mathcal{L}}}
\newcommand{\nn}{\nonumber \\}
\newcommand{\e}{\mathrm{e}}
\newcommand{\tr}{\mathrm{tr}\,}

\tolerance=5000

\title{Entropy for curvature squared gravity using surface term and auxiliary field}

\author{Kyosuke Hirochi$^1$}

\affiliation{
$^1$ Department of Physics, Nagoya University, Nagoya
464-8602, Japan 
}

\def\theequation{\thesection.\arabic{equation}}



\begin{abstract}
We compute the entropies for general curvature squared gravities in arbitrary dimensions using the conserved charge and Virasoro algebra from surface term.
We introduce an auxiliary tensor field in order to obtain the boundary action which establish a variational principle. Our results reproduce those in some specific models, that is, the Gauss-Bonnet, new massive gravity (NMG), and critical gravity.
\end{abstract}


\maketitle


\section{Introduction}

Black hole is an important object in the theory of gravity because studying the properties of the black hole can give us to some indications for the quantum gravity.
A thermal behavior is one of the remarkable aspect of black hole \cite {Bekenstein:1973ur,Bardeen:1973gs}. 
In the concept of black hole thermodynamics for the Einstein gravity, temperature and entropy correspond to surface gravity and area of the black hole horizon, respectively.
The methods to obtain the entropy have developed in last few decades. 
One of them is to use the Virasoro algebra with central charge and the Cardy formula \cite{Carlip:1998wz, Carlip:1999cy}.
In this method, the conserved charge constructing this algebra is related with the diffeomorphism and obtained from the bulk action.

Recently, a new method was proposed in \cite{sur1,Majhi:2012nq}.
In Refs papers, \cite{sur1,Majhi:2012nq}, the Virasoro algebra and the conserved charge are obtained from the surface term instead of the bulk term.
We can obtain the central charge and the zero mode eigenvalues of the Fourier modes of the charge from the Virasoro algebra without any ambiguity like shifting the zero mode charge and the choice of parameter $\alpha$ appeared in Fourier modes of T which is determined as the solution of the Killing equations (detail is written in section 2).
Using the Cardy formula \cite{Cardy:1986ie}, we obtain the black hole entropy.
In the Einstein gravity, the corresponding surface term is the well-known Gibbons-Hawking term, and reproduces the Bekenstein-Hawking entropy.

Our purpose in this paper is application of the method to the general curvature squared gravity. 
For the special case including higher curvature terms, the Gauss-Bonnet gravity, the corresponding entropy was calculated in \cite{Zhang:2012fq} (in this paper the third order Lovelock model was also calculated).
However, for generic gravitational theories including curvature squared term we have not obtained entropy because there is no corresponding surface term that leads to a well-posed variational principle.  
This problem can be improved by introducing an auxiliary field which rewrite the action as a second-order form.
Due to this reduced action, we can obtain a generalized Gibbons-Hawking term without loosing the variational principle.

 This paper is organized as follows. 
In the next section, we will review how to obtain the conserved Noether charge from the surface term of a general gravitational action.
In section 3, we will obtain the surface term for curvature squared action using the method of introducing auxiliary tensor field. 
In section 4, we will compute the generic formula of entropy for the Schwarzschild black hole.
We will conclude our result in last section.

\section{Noether currents and their charges from surface term}
In this section, we review the calculation Noether current and conserved charge using the surface term.
In general, the surface term of the gravitational action on the manifold {\it M} in {\it D}-dimensions is given by
\begin{eqnarray}
S_{\rm B}=\frac{1}{16\pi G}\int_{\partial {\it M}} d^{{\it D}-1}x\sqrt{-\gamma}{\cal L}_{\rm B}
=\frac{1}{16\pi G}\int_{{\it M}}d^{\it D}x\sqrt{-g}\nabla_\mu(N^\mu{\cal L}_{\rm B}),
\end{eqnarray}
where $N^\mu$ is the unit normal of the boundary $\partial {\it M}$, $g_{\mu\nu}$ is the metric on the {\it M }and $\gamma_{\mu\nu}$ is the boundary metric induced from $g_{\mu\nu}$.
The conserved Noether current $J^\mu$ which we like to obtain corresponds to a general diffeomorphism transformation
\begin{equation}
x^\mu \rightarrow x^\mu+\xi^\mu.\label{diff}
\end{equation}
Now, we define the Lagrangian density for convenience,
\begin{equation}
\sqrt{-g}{\cal L}=\sqrt{-g}\nabla_\mu A^\mu,\quad A^\mu\equiv \frac{1}{16\pi G}N^\mu{\cal L}_{\rm B},\label{Lde}
\end{equation}
where ${\cal L}$ is a scalar. Under (\ref {diff}), the left hand side of (\ref{Lde}) varies by
\begin{equation}
\delta_{\xi}(\sqrt{-g}{\cal L})\equiv\pounds_{\xi}(\sqrt{-g}{\cal L})=\sqrt{-g}\nabla_\mu({\cal L}\xi^\mu).\label{left}
\end{equation}
On the other hand, the right hand side changes by
\begin{eqnarray}
\delta_{\xi}(\sqrt{-g}\nabla_\mu A^\mu)&=&\pounds_\xi\left[\partial_\mu(\sqrt{-g}A^\mu)\right]
=\partial_\mu\left[\pounds_\xi\sqrt{-g}A^\mu+\sqrt{-g}\pounds_\xi A^\mu\right]
=\sqrt{-g}\nabla_\mu\left[\nabla_\nu(A^\mu\xi^\nu)-A^\nu\nabla_\nu\xi^\mu\right],\label{right}
\end{eqnarray}
where we have used the relationships
\begin{equation}
\nabla_\mu A^\mu=\frac{1}{\sqrt{-g}}\partial_\mu(\sqrt{-g}A^\mu).
\end{equation}
Then, we get the conserved Noether current by equating (\ref{left}) and (\ref{right}),
\begin{equation}
J^\mu[\xi]={\cal L}\xi^\mu-\nabla_\nu(A^\mu\xi^\nu)+A^\nu\nabla_\nu\xi^\mu,\quad \nabla_{\mu}J^\mu=0.
\end{equation}
Also we can define the Noether potential $J^{\mu\nu}$ by rewriting $J^\mu$ as
\begin{equation}
J^{\mu}[\xi]=\nabla_{\nu}J^{\mu\nu}[\xi]=\nabla_\nu(\xi^\mu A^\nu-\xi^\nu A^\mu).
\end{equation}
The corresponding charge is then given by
\begin{equation}
Q[\xi]=\frac{1}{2}\int_{\rm M}\sqrt{-g}d\Sigma_\mu J^\mu=\frac{1}{2}\int_{\rm M}\sqrt{-g}d\Sigma_\mu \nabla_\nu J^{\mu\nu}=\frac{1}{2}\int_{\partial M}\sqrt{-h}d\Sigma_{\mu\nu}J^{\mu\nu},\label{cha}
\end{equation}
where the last equality is coming from the Stoke's theorem and $d\Sigma_{\mu\nu}$ is the ({\it D}-2)-dimensional surface element defined as
\begin{equation}
d\Sigma_{\mu\nu}\equiv-d^{{\it D}-2}x(N_\mu M_\nu-N_\nu M_\mu),
\end{equation}
and $h$ is the determinant of the corresponding induced metric. 
Here, $N^\mu$ and $M^\mu$ are chosen as the space-like and time-like unit normals respectively:
\begin{equation}
N^\mu N_\mu=1,\quad M^\mu M_\mu=-1,\quad N^\mu M_\mu=0.
\end{equation}

Next, we define the algebra of the Noether charges (\ref{cha}) as
\begin{equation}
[Q_1,Q_2]\equiv(\delta_{\xi_1} Q[\xi_2]-\delta_{\xi_2} Q[\xi_1]).\label{al}
\end{equation}
We will show that this algebra leads to the Virasoro algebra with the central charge. First, we compute the variation of the Noether charge
\begin{equation}
\delta_\xi Q[\xi]=\int d\Sigma_\mu \delta_\xi(\sqrt{-g}J^\mu)=\int d\Sigma_{\mu\nu}\sqrt{-h}\xi^\nu J^\mu.
\end{equation}
So, the algebra (\ref{al}) becomes
\begin{equation}
[Q_1,Q_2]=\int d \Sigma_{\mu\nu}\sqrt{-h}\left(\xi_2^\mu J^\nu[\xi_1]-\xi_1^\mu J^\nu[\xi_2]\right).
\end{equation}

Now, we consider the general static spherical black holes solutions whose metric is given by
\begin{equation}
ds^2=-f(r)dt^2+f(r)^{-1}dr^2+h_{ij}dx^idx^j.\label{sw}
\end{equation}
Here $f(r)$ is a function which satisfies the conditon $f(r_h)=0$, where $r=r_h$ is the location of the horizon.
From now on, we focus on the near horizon regions, so it is convenient to define a new coordinate $\rho$
\begin{equation}
r\equiv r_h+\rho,\quad\rho\ll 1.
\end{equation}
In these regions, the function $f(r_h+\rho)$ can be expanded as
\begin{equation}
f(r_h+\rho)=0+f'(r_h)\rho+\frac{1}{2}f^{''}(r_h)\rho^2+\cdots.
\end{equation}
Here the surface gravity is determined by
\begin{equation}
\kappa\equiv\frac{f'(r_h)}{2}.
\end{equation}
On the horizon, $\rho\rightarrow 0$, the metric becomes singular, so we introduce the Bondi-like coordinates defined by
\begin{equation}
du=dt-f(\rho)^{-1}d\rho.
\end{equation}
Thus the metric (\ref{sw}) becomes regular on the horizon,
\begin{equation}
ds^2=-f(\rho)dt^2-2dud\rho+h_{ij}dx^idx^j.
\end{equation}
We solve the Killing equations for the vector field $\xi^\mu$ which leave the horizon structure invariant
\begin{eqnarray}
0=\pounds_\xi g_{\rho\rho}=-2\partial_\rho\xi^u,\quad
0=\pounds_\xi g_{u\rho}=-f(\rho)\partial_\rho\xi^u-\partial_\rho\xi^\rho-\partial_u\xi^u,\label{rhou}
\end{eqnarray}
\begin{equation}
0=\pounds_\xi g_{uu}=-\xi^\rho\partial_\rho f(\rho)-2f(\rho)\partial_u\xi^u-2\partial_u\xi^\rho={\cal O}(\rho)\label{uu}.
\end{equation}
We should note that the equation (\ref{uu}) near the horizon is trivially satisfied. 
The solutions for the Killing equations (\ref{rhou}) are given by
\begin{equation}
\xi^u=F(u,x),\quad\xi^\rho=-\rho\partial_t F(u,x),
\end{equation}
where $F(u,x)$ is an arbitrary function and other components of $\xi^\mu$ vanish. 
In the original coordinates $(t,\rho)$, these Killing vectors have the following forms
\begin{equation}
\xi^t=T-\frac{\rho}{f(\rho)}\partial_t T,\quad\xi^\rho=-\rho\partial_t T,\quad T(t,\rho,x)\equiv F(u,x).
\end{equation}
We define the two normal vectors $N^\mu,M^\mu$ for the metric (\ref{sw}) as follows,
\begin{equation}
N^\mu=(0,\sqrt{f(\rho)},0,\cdots,0),\quad M^\mu=\left(\frac{1}{\sqrt{f(\rho)}},0,\cdots,0\right)\label{nm}.
\end{equation}
Substituting (\ref{nm}) into (\ref{cha}), we obtained the conserved Noether charge,
\begin{equation}
Q[\xi]=\frac{1}{16\pi G}\int\sqrt{-h}d^{{\it D}-2}x\frac{\rho}{\sqrt{f}}\left(\frac{f}{\rho}T-\partial_t T\right){\cal L}_{\rm B}.
\end{equation}

\section{Surface term for curvature squared term}

In order to evaluate the entropy, we need to calculate the surface term ${\cal L}_{\rm B}$. 
In general, however, ${\cal L}_{\rm B}$ for the higher-derivative term cannot be obtained because we impose the boundary condition that the variations of the metric should vanish, higher derivative terms of the variations are still exist, so there is no variational principle except with some specific term (for examples, the Gauss-Bonnet and the Lovelock gravities).
As a substituted method, we use auxiliary field method to get the action including up to second-derivative term which is equivalence with the curvature squared one on-shell \cite{Hohm:2010jc} (see also \cite{Nojiri:2001ae} as a pioneer work ).

We consider the most general curvature squared pure gravity action in {\it D}-dimensions as
\begin{equation}
S=\frac{1}{16\pi G}\int d^Dx\sqrt{-g}\left(\sigma R-2\Lambda_0+a_1R^{\mu\nu\rho\sigma}R_{\mu\nu\rho\sigma}+a_2R^{\mu\nu}R_{\mu\nu}+a_3R^2\right), \label{ac}
\end{equation}
where $\Lambda_0$ is the bare cosmological constant and $\sigma$ takes the values of $0$ or $\pm 1$  and $a_1,$ $a_2$, and $a_3$ are the coupling constants.
For the Einstein-Hilbert term, the surface term is well-known so called hGibbons-Hawking termh
\begin{equation}
S_{\rm GH}=\frac{1}{16\pi G}\int d^{D-1}x\sqrt{-\gamma}(-2\sigma K),
\end{equation}
where $K$ is the trace of extrinsic curvature defined by,
\begin{equation}
K_{\mu\nu}\equiv-\frac{1}{2}(\nabla_\mu N_\nu+\nabla_\nu N_\mu).
\end{equation}
 
We restrict to consider the action including only second-order derivative terms by introducing an auxiliary field $\phi_{\mu\nu\rho\sigma}$,

\begin{equation}
S_\phi=\frac{1}{16\pi G}\int d^Dx\sqrt{-g}(\phi^{\mu\nu\rho\sigma}R_{\mu\nu\rho\sigma}+b_1\phi^{\mu\nu\rho\sigma}\phi_{\mu\nu\rho\sigma}+b_2\phi^{\mu\nu}\phi_{\mu\nu}+b_3\phi^2)\label{sphi},
\end{equation}
where $\phi_{\mu\nu\rho\sigma}$ has the same symmetries as the Riemann tensor $\phi_{\mu\nu\rho\sigma}=-\phi_{\mu\nu\sigma\rho}=-\phi_{\nu\mu\rho\sigma},\quad \phi_{\mu\nu\rho\sigma}=\phi_{\rho\sigma\mu\nu}$ etc. and $\phi_{\mu\nu}\equiv g^{\rho\sigma}\phi_{\rho\mu\sigma\nu},\quad \phi\equiv g^{\mu\nu}\phi_{\mu\nu}$.
We show that by choosing the parameters $(b_1,b_2,b_3)$ properly, $S_{\phi}$ becomes equivalent to the curvature squared term of (\ref{ac}).
The equation for $\phi$ is given by
\begin{equation}
R_{\mu\nu\rho\sigma}+2b_1\phi_{\mu\nu\rho\sigma}+2b_2\phi_{\langle\mu\rho}g_{\nu\sigma\rangle}+2b_3\phi g_{\langle\mu\rho}g_{\nu\sigma\rangle}=0,
\label{phi}\end{equation}
where $\langle\cdots\rangle$ expresses, for example,
\begin{equation}
\phi_{\langle\mu\rho}g_{\nu\sigma\rangle}=\frac{1}{4}(\phi_{\mu\rho} g_{\nu\sigma}-\phi_{\nu\rho}g_{\mu\sigma}-\phi_{\mu\sigma}g_{\nu\rho}+\phi_{\nu\sigma}g_{\mu\rho}).
\end{equation}
Taking the trace of (\ref{phi}), one obtains
\begin{equation}
R_{\mu\nu}+2b_1\phi_{\mu\nu}+\frac{b_2}{2}\left[(D-2)\phi_{\mu\nu}+\phi g_{\mu\nu}\right]+(D-1)b_3\phi g_{\mu\nu}=0,
\end{equation}
and,
\begin{equation}
R+\left[2b_1+(D-1)b_2+D(D-1)b_3\right]\phi=0.
\end{equation}
Then, we obtain the relations between $\phi_{\mu\nu\rho\sigma}$ and $R_{\mu\nu\rho\sigma}$ as follows
\begin{eqnarray}
\phi=-\frac{R}{2b_1+(D-1)b_2+D(D-1)b_3},
\end{eqnarray}
\begin{equation}
 \phi_{\mu\nu}=-\frac{2}{4b_1+(D-2)b_2}\left(R_{\mu\nu}-\frac{b_2+D(D-1)b_3}{2[2b_1+(D-1)b_2+D(D-1)b_3]}Rg_{\mu\nu}\right), \\ 
\end{equation}
\begin{eqnarray}
\phi_{\mu\nu\rho\sigma}&=&-\frac{1}{2b_1}\Big(R_{\mu\nu\rho\sigma}-\frac{4b_2}{4b_1+(D-2)b_2}R_{\langle\mu\rho}g_{\nu\sigma\rangle}\\
&&+\frac{2[b^2_2-4b_1b_3+Db_2b_3]}{[4b_1+(D-2)b_2][2b_1+(D-1)b_2+D(D-1)b_3]}Rg_{\langle\mu\rho}g_{\nu\sigma\rangle}\Big).\nonumber
\end{eqnarray}
Inserting these expressions into (\ref{sphi}) and compared with (\ref{ac}), we finally obtain the relationships between the ($a_1,a_2,a_3$) and ($b_1,b_2,b_3$) as follows
\begin{eqnarray}
a_1&=&-\frac{1}{4b_1},\\
a_2&=&\frac{b_2}{b_1[4b_1+(D-2)b_2]},\\
a_3&=&-\frac{b^2_2-4b_1b_3+Db_2b_3}{2b_1[4b_1+(D-2)b_2][2b_1+(D-1)b_2+D(D-1)b_3]}.
\end{eqnarray}
Next, in order to obtain the general Gibbons-Hawking (GGH) term we perform the variation of (\ref{sphi}). Since only a surface term we interested in, we ignore a term which vanishes on-shell,
\begin{eqnarray}
\delta_{g} S_{\phi}&=&\frac{1}{16\pi G}\int d^Dx\sqrt{-g}\phi^{\mu\nu\rho\sigma}\delta R_{\mu\nu\rho\sigma}\nonumber\\
&=&\frac{1}{16\pi G}\int d^Dx\sqrt{-g}\nabla_\mu\left(-2\phi^{\mu\nu\rho\sigma}\nabla_\rho\delta g_{\nu\sigma}+2\nabla_\rho\phi^{\mu\nu\rho\sigma}\delta g_{\nu\sigma}\right)\nonumber\\
&=&\frac{1}{16\pi G}\int d^{D-1}x\sqrt{-\gamma}N_\mu(-2\phi^{\mu\nu\rho\sigma}\nabla_\rho\delta g_{\nu\sigma}),
\end{eqnarray}
where we imposed the boundary condition $\delta g_{\mu\nu}=0$ to the third line. 
Using the variations of $K_{\mu\nu}$ on surface, the integrand of $\delta_{g} S_{\phi}$ becomes
\begin{equation}
-2N_\mu\phi^{\mu\nu\rho\sigma}\nabla_\rho\delta g_{\nu\sigma}=4N_r\phi^{rirj}\delta K_{ij},
\end{equation}
where component $r$ expresses the direction normal to the boundary $\partial M$. 
Then, we get the generalized Gibbons-Hawking term as follows
\begin{equation}
S_{\rm GGH}=-\frac{1}{16\pi G}\int d^{D-1}x\sqrt{-\gamma}(2\sigma+4\phi^{rirj}K_{ij}).
\end{equation}

Using the (A)dS backgrounds solutions of (\ref{ac}) where
\begin{eqnarray}
R_{\mu\nu\rho\sigma}=\frac{2\Lambda}{(D-1)(D-2)}(g_{\mu\rho}g_{\nu\sigma}-g_{\mu\sigma}g_{\nu\rho}),\quad
R_{\mu\nu}=\frac{2\Lambda}{D-2}g_{\mu\nu},\quad R=\frac{2D\Lambda}{D-2},
\end{eqnarray}
\begin{equation}
\Lambda_0=\sigma\Lambda+\frac{2(D-4)}{D-2}\left(\left[D(a_3-a_1)+a_2+4a_1\right]\frac{1}{D-2}+a_1\frac{D-3}{D-1}\right)\Lambda^2,
\end{equation}
we obtain the $S_{\rm GGH}$ for the (A)dS background
\begin{equation}
S_{\rm GGH}=-\frac{1}{8\pi G}\int d^{D-1}x\sqrt{-\gamma}\left[(\sigma+F)K\right],
\end{equation}
where F is defined as the contribution to the ${S_{\rm GH}}$,
\begin{equation}
F\equiv\frac{4\Lambda}{(D-1)(D-2)}\left[2a_1+(D-1)(a_2+Da_3)\right].
\end{equation}

\section{Entropy for curvature squared gravity}
In the previous section, the GGH action for the auxiliary field methods is obtained which is equivalent with the original curvature squared action on-shell.
Now, we are ready to calculate the entropy for the general curvature squared gravity. 
For the Schwarzschild type metric, discussed in section 2, the extrinsic curvature is rewritten by 
\begin{equation}
K=-\frac{1}{2\sqrt{f}}\partial_\rho f,
\end{equation}
and the conserved charge near the horizon becomes
\begin{equation}
Q[\xi]=\frac{1}{8\pi G}\int d^{D-2}x\sqrt{-h}\left(\kappa T-\frac{1}{2}\partial_t T\right)(\sigma+F).\label{q}
\end{equation}
We expand $T$ in term of basis function $T_m$
\begin{equation}
T=\sum_m A_m T_m,\quad A_m^*=A_{-m},
\end{equation}
\begin{equation}
T_m=\frac{1}{\alpha}{\rm exp}\left[im(\alpha t+g(\rho)+p\cdot x)\right]\label{tm}.
\end{equation}
where $\alpha$ is a arbitrary parameter, $p$ is an integer, and $g(\rho)$ is a function that is regular at the horizon.
Here $T_m$ is chosen to obey the algebra isometric to Diff $S^1$ \cite{Silva:2002jq}
\begin{equation}
i\{\xi_m,\xi_n\}^a=(m-n)\xi^a_{m+n}.
\end{equation}
Substituting (\ref{tm}) to (\ref{q}) and we obtain $Q_m$ determined by $Q=\sum_m Q_mA_m$ and their Virasoro algebra
\begin{equation}
Q_m=\frac{\kappa A}{8\pi G\alpha}\left(\sigma+F\right)\delta_{m,0},
\end{equation}
\begin{equation}
[Q_m,Q_n]=\frac{i\kappa A}{8\pi G\alpha}\left(\sigma+F\right)(m-n)\delta_{m+n,0}
-\frac{im^3\alpha A}{16\pi G\kappa}\left(\sigma+F\right)\delta_{m+n,0}.
\end{equation}
From the Virasoro algebra and the charge, we can get the central charge c and the zero mode $Q_0$ as
\begin{equation}
\frac{c}{12}=\frac{\alpha}{16\pi G\kappa}A\left(\sigma+F\right),\quad Q_0=\frac{\kappa}{8\pi G\alpha}A\left(\sigma+F\right),
\end{equation}
where $A$ is the surface area of black hole.
 Using the Cardy formula, we obtain the entropy of the curvature squared gravity,
 \begin{equation}
S=2\pi\sqrt{\frac{cQ_0}{6}}=\frac{A}{4}(\sigma +F).
\end{equation}
We show several specific examples. First, the Einstein+Gauss-Bonnet type gravity, $(a_1,a_2,a_3)=(1,-4,1)$, the entropy becomes
\begin{equation}
S=A\left[\sigma+\frac{4a(D-3)}{(D-1)}\Lambda\right].
\end{equation}
where $a$ is a coefficient of the Gauss-Bonnet term.
This result coincides with that in \cite{Cai:2001dz}. 
 Second, the new massive gravity (NMG) in 3-Dim \cite{Bergshoeff:2009hq}, $(a_1,a_2,a_3)=(0,-1,\frac{3}{8})$,
\begin{equation}
S=\left(\sigma-\frac{1}{2m^2}\Lambda\right),
\end{equation}
where $m^2$ is a coefficient of the higher curvature term.
Finally we discuss Critical Gravity in 4-Dim,$(a_1,a_2,a_3)=(0,-\frac{3}{2\Lambda},-\frac{1}{2\Lambda})$.
In this parameter choice, the entropy vanishes, which is same as \cite{Lu:2011zk}.

\section{Conclusion}
In this paper we have studied the entropy of $D$-dimensional gravity with curvature squared term 
using the method based on the Virasoro algebra and the central charge. 
Introducing an auxiliary field we have obtained the second-derivative formed action which is equivalent with the original one on-shell.
Thus, without loosing the variational principle, we have calculated the surface action and the Black Hole entropy.

We note that the above technique is valid for the Schwarzschild type metric. 
So, the study whether the same method is still make sure to the other types of metric, the Kerr metric for example, is the interesting future works.


\begin{thebibliography}{20}
 
\bibitem{Bekenstein:1973ur} 
  J.~D.~Bekenstein,
  ``Black holes and entropy,''
  Phys.\ Rev.\ D {\bf 7}, 2333 (1973).
\bibitem{Bardeen:1973gs} 
  J.~M.~Bardeen, B.~Carter and S.~W.~Hawking,
  ``The Four laws of black hole mechanics,''
  Commun.\ Math.\ Phys.\  {\bf 31}, 161 (1973).
 

\bibitem{Carlip:1998wz} 
  S.~Carlip,
  ``Black hole entropy from conformal field theory in any dimension,''
  Phys.\ Rev.\ Lett.\  {\bf 82}, 2828 (1999)
  [hep-th/9812013].
 
\bibitem{Carlip:1999cy} 
  S.~Carlip,
  ``Entropy from conformal field theory at Killing horizons,''
  Class.\ Quant.\ Grav.\  {\bf 16}, 3327 (1999)
  [gr-qc/9906126].
 
 \bibitem{sur1}
 B.~R.~Majhi and T.~Padmanabhan,
  ``Noether current from the surface term of gravitational action, Virasoro algebra and horizon entropy,''
  Phys.\ Rev.\ D {\bf 86}, 101501 (2012)
  [arXiv:1204.1422 [gr-qc]].

\bibitem{Majhi:2012nq} 
  B.~R.~Majhi,
  ``Noether current of the surface term of Einstein-Hilbert action, Virasoro algebra and entropy,''
  arXiv:1210.6736 [gr-qc].



\bibitem{Cardy:1986ie} 
  J.~L.~Cardy,
  ``Operator Content of Two-Dimensional Conformally Invariant Theories,''
  Nucl.\ Phys.\ B {\bf 270}, 186 (1986).

\bibitem{Zhang:2012fq} 
  S.~-J.~Zhang and B.~Wang,
  ``Surface term, Virasoro algebra and Wald entropy of black holes in higher curvature gravity,''
  Phys.\ Rev.\ D {\bf 87}, 044041 (2013)
  [arXiv:1212.6896 [hep-th]].


\bibitem{Hohm:2010jc} 
  O.~Hohm and E.~Tonni,
  ``A boundary stress tensor for higher-derivative gravity in AdS and Lifshitz backgrounds,''
  JHEP {\bf 1004}, 093 (2010)
  [arXiv:1001.3598 [hep-th]].

\bibitem{Nojiri:2001ae} 
  S.~Nojiri, S.~D.~Odintsov and S.~Ogushi,
  ``Cosmological and black hole brane world universes in higher derivative gravity,''
  Phys.\ Rev.\ D {\bf 65}, 023521 (2001)
  [hep-th/0108172].



\bibitem{Silva:2002jq} 
  S.~Silva,
  ``Black hole entropy and thermodynamics from symmetries,''
  Class.\ Quant.\ Grav.\  {\bf 19}, 3947 (2002)
  [hep-th/0204179].

\bibitem{Cai:2001dz} 
  R.~-G.~Cai,
  ``Gauss-Bonnet black holes in AdS spaces,''
  Phys.\ Rev.\ D {\bf 65}, 084014 (2002)
  [hep-th/0109133].

\bibitem{Bergshoeff:2009hq} 
  E.~A.~Bergshoeff, O.~Hohm and P.~K.~Townsend,
  ``Massive Gravity in Three Dimensions,''
  Phys.\ Rev.\ Lett.\  {\bf 102}, 201301 (2009)
  [arXiv:0901.1766 [hep-th]].

\bibitem{Lu:2011zk} 
  H.~Lu and C.~N.~Pope,
  ``Critical Gravity in Four Dimensions,''
  Phys.\ Rev.\ Lett.\  {\bf 106}, 181302 (2011)
  [arXiv:1101.1971 [hep-th]].
\end{thebibliography}
\end{document}